\documentclass[twocolumn,showpacs,amsmath,amssymb,pra,footinbib,floatfix,superscriptaddress]{revtex4}

\usepackage{graphicx}
\usepackage{dcolumn}
\usepackage{bm}
\usepackage{helvet}
\usepackage{amssymb}
\usepackage{amsmath}
\usepackage{amsfonts}
\usepackage{hyperref}
\usepackage{color}
\usepackage{epstopdf}

\newcommand{\beq}{\begin{equation}}
\newcommand{\eeq}{\end{equation}}

\usepackage{calc}

\date{\today}
\begin{document}

\author{Cristian Degli Esposti Boschi}
\address{CNR-IMM, Sezione di Bologna, via Gobetti 101, I-40129, Bologna, Italy}

\author{Elisa Ercolessi}
\address{Dipartimento di Fisica e Astronomia dell'Universit\`a di Bologna, Via Irnerio 46, 40127 Bologna, Italy}
\address{INFN, Sezione di Bologna, Via Irnerio 46, 40127 Bologna, Italy}

\author{Loris Ferrari}
\address{Dipartimento di Fisica e Astronomia dell'Universit\`a di Bologna, Via Irnerio 46, 40127 Bologna, Italy}

\author{Piero Naldesi}
\address{Dipartimento di Fisica e Astronomia dell'Universit\`a di Bologna, Via Irnerio 46, 40127 Bologna, Italy}
\address{INFN, Sezione di Bologna, Via Irnerio 46, 40127 Bologna, Italy}

\author{Fabio Ortolani}
\address{Dipartimento di Fisica e Astronomia dell'Universit\`a di Bologna, Via Irnerio 46, 40127 Bologna, Italy}
\address{INFN, Sezione di Bologna, Via Irnerio 46, 40127 Bologna, Italy}

\author{Luca Taddia}
\email{luca.taddia2@gmail.com}
\address{Scuola Normale Superiore, Piazza dei Cavalieri 7, 56126 Pisa, Italy}
\address{CNR - Istituto Nazionale di Ottica, UOS di Firenze LENS, Via Carrara 1, 50019 Sesto Fiorentino, Italy}

\title{Bound states and expansion dynamics of interacting bosons on a one-dimensional lattice}

\begin{abstract}

The expansion dynamics of bosonic gases in optical lattices has recently been the focus of an incresing attention, both experimental and theoretical. We consider, by means of numerical Bethe ansatz, the expansion dynamics of initially confined wave packets of two interacting bosons on a lattice. We show that a correspondence between the asymptotic expansion velocities and the projection of the evolved wave function over the bound states of the system exists, clarifying the existing picture for such situations. Moreover, we investigate the role of the lattice in this kind of evolution.

\end{abstract}


\maketitle



\section{Introduction}\label{Intro}

Since the first experimental realizations of Bose-Einstein condensates \cite{BEC_exp}, huge efforts have been devoted to the study of ultra-cold bosonic atoms loaded into magnetic-optical traps \cite{Reviews}. In these experiments the density profile after release from the magnetic trap represents a key quantity. In the hypothesis of a ballistic, i.e., free, expansion, the density profile can be easily related to the momentum distribution of the initial quantum state.

In most situations, interactions play a key role and  cannot be neglected during the expansion. In particular, for one-dimensional (1D) systems, this leads to a variety of new interesting phenomena that might be predicted and observed: the non-thermalization of a Lieb-Liniger gas \cite{Kinoshita2006}, the dynamical fermionization of expanding hard-core bosons \cite{DynFerm}, the quantum distillation of double occupations \cite{HM2009,Ronzheimer2013,Bolech2012} and the dynamical quasi-condensation of an initial product state \cite{QuasiCond}. In particular, the latter consists in the observation of the formation, during the evolution, of two quasi-coherent matter beams and a momentum distribution function peaked around two opposite values, a very different behaviour from the one expected in the continuum case \cite{Messiah1999}. It is worth mentioning that expansion is a possible playground for the study of  dynamical properties of quantum many body models, such as, e.g., the dynamics of entanglement \cite{Kessler2013}. In the last years a lot of attention has been devoted to these topics (see Ref. \onlinecite{Polkovnikov2011} for a review), among which we may recall the problem of thermalization (or its absence) \cite{Kinoshita2006,Kormos2009} and the formation of topological defects after the crossing of a critical point by means of the Kibble-Zurek mechanism \cite{Kibble,Zurek,Lamporesi2013,Zurek2005,Braun2014,Canovi2014}.

One of the most  striking peculiarities of 1D many body-models is the existence of a large class of potentials that admit two-(pseudo)particle bound states. This is the case both in the continuum, e.g., in the Lieb-Liniger-Yang model \cite{LLY}, and for lattice models, such as spin models, where magnons can bound as predicted by H. A. Bethe more than eighty years ago \cite{Bethe}, or interacting fermionic systems, as pointed out by J. Hubbard himself in his seminal work \cite{Hubbard}. Most interestingly, as a consequence of the discrete nature of the lattice, bound states might exist not only for the attractive case, but also for repulsive interactions, when they can become stable in absence of dissipation, as it is the case when the system is integrable \cite{Sutherland}. In recent times, the possibility of observing bound states in a cold-atoms setup by means of dynamical probes has drawn a lot of attention, both theoretically \cite{Ganahl2012} and experimentally \cite{Fukuhara2013}. The existence of a stable bound state is also the key ingredient for the phenomenon of induced resonances which are at the origin of the BEC-BCS crossover effect in atomic systems \cite{Pethick}, in Fermi gases \cite{Zwerger} as well for polar molecules \cite{dipolar}. Very recently, bound states have also entered the debate about equilibration and thermalization \cite{Goldstein2014}.

For bosonic systems, the existence of bound states in lattice models has been investigated theoretically and experimentally, both from a static \cite{Valiente,Javanainen2010,Winkler2006} and a dynamic \cite{Ganahl2012, Fukuhara2013} point of view. They can, in principle, play an important role in the expanding dynamics of a lattice Bose gas: it is natural to expect that if the wave function of the system possesses a large projection over the set of the bound states, the dynamics should be quasi-stationary, and the expansion very different from a ballistic one. This projection is hard to study since, for a non-integrable model such as Bose-Hubbard, the explicit form of the bound-state eigenfunctions are in general analytically not known. Still, the two-body problem might be exactly solvable and therefore the study of the role of the bound states should be, in that case, possible, in order to get some hint about the many-body case. 

In this article, we are going to investigate the dynamics of two expanding bosons on a lattice, as ruled by the Hamiltonian of the Bose-Hubbard model \cite{Fisher1989}:
\begin{eqnarray}\label{BHH}
 	\hat{H}(U)&=&-J\sum_{j=-L/2}^{L/2}\left(b_j^\dagger b_{j+1}+b_{j+1}^\dagger b_j\right)+\nonumber\\
        &+&\frac{U}{2}\sum_{j=-L/2}^{L/2}\hat{n}_j\left(\hat{n}_j-1\right)
\end{eqnarray}
where $J$ and $U$ are respectively the hopping and the on-site-interaction coefficients, $b_j$ is a bosonic annihilation operator and  $\hat{n}_j=b^\dagger_j b_j$.
Our work is motivated by the experimental results of Ref. \onlinecite{Ronzheimer2013}, where the authors studied the expansion by means of the Hamiltonian (\ref{BHH}) of an initial product state in real space, which, for $U/J<+\infty$, was not an eigenstate of $\hat{H}$ itself. Interestingly, they found that the dynamics in both the free ($U=0$) and the hard-core ($U=+\infty$) case displays the same {\it expansion velocity}, defined as
\begin{equation}\label{expv}
     v=\frac{d}{dt}\sqrt{R^2(t)-R^2(0)}
\end{equation}
where $R^2(t)$ represents the time-dependent variance of the density distribution:
\begin{equation}\label{Rsquare}
     R^2(t)=\frac{a^2}{N}\sum_{j=-L/2}^{L/2}n_j(t)\left(j-j_0\right)^2
\end{equation}
$a$ and $j_0$ being the lattice spacing and the centre of the initial wave packet. Moreover, the expansion velocity has a minimum at an intermediate value of $U$.
By analyzing a number of different situations, we will show that, independently from the kind of initial state, there exists a strong relation between the expansion velocity and the presence of bound states in the spectrum. In addition, we discuss a number of quantitative and qualitative features about the expansion of initially confined bosons on a 1D lattice, as well as the role of the lattice during the expansion, compared to the continuum case.

This article is organized as follows. We derive in  Sec. \ref{BAS} the exact solution of the two-bosons problem for the Hamiltonian  \eqref{BHH} on a finite lattice with periodic boundary conditions. We find scattering and bound-state eigenfunctions, and show that the latter exist for both attractive and repulsive interactions. We then analyze the effects of bound states in the dynamical evolution of the system, when the initial state is not an eigenstate of the Hamiltonian. We consider two cases: i) in Sec. \ref{product} the initial state is chosen to be a product state in real space, in which the two bosons are put in a single or in two neighbor sites; ii) in Sec. \ref{entangled}  we consider an initial superfluid-like two-particle wave packet,  which is non-factorizable in real space, as it might be experimentally feasible in a cold-boson setup \cite{Greiner2002}. We collect some results about the expansion of a single boson on a lattice in Sec. \ref{lattice}. In Sec. \ref{conclusions} we draw our conclusions and comment on possible developments. Finally in Appendices \ref{freevel} and \ref{theorem} we respectively discuss the formulas for the expansion velocities in the non-interacting case and recall the scenario of the $U\leftrightarrow-U$ inversion theorem, as formulated in Ref. \onlinecite{Schneider2012}.


\section{The two-boson problem on a lattice: exact results}\label{BAS}


In this and in the following Sections we are going to consider the Bose-Hubbard Hamiltonian \eqref{BHH} for  $N=2$ particles. In this case, despite the non-integrability of the model \cite{Cazalilla2011}, it is possible to solve the Schr\"odinger equation exactly, by the Bethe ansatz technique, thanks to the separation of centre-of-mass and relative coordinates. A similar analysis was exploited by Valiente and Petrosyan in \cite{Valiente} in the thermodynamic limit; instead, we work at {\it finite} size $L+1$, with $L$ even. Moreover, in the rest of the paper (unless otherwise stated), we will choose $\hbar=a=1$. 

The eigenstates of $\hat{H}$ take the form
 \begin{equation}\label{psi}
 	\left|\phi\right>=\sum_{j,k=-L/2}^{L/2}\phi_{jk}b_j^\dagger b_k^\dagger\left|0\right>
 \end{equation}
where the coefficients $\phi_{jk}$, symmetric under the exchange of $j$ and $k$ and properly normalized, satisfy the equations:
 \begin{eqnarray}
 	&&J\left(\phi_{j+1,k}+\phi_{j-1,k}+\phi_{j,k+1}+\phi_{j,k-1}\right)+\nonumber\\
        &&-\left(U\delta_{jk}-E\right)\phi_{jk}=0\label{EShIT}
 \end{eqnarray}

We now look for solutions of the Bethe Ansatz form  \cite{Takahashi1999}:
\begin{eqnarray}\label{BA}
    \phi_{jk}&=&\left[a_{12}e^{i\left(p_1j+p_2k\right)}+a_{21}e^{i\left(p_1k+p_2j\right)}\right]\vartheta(j-k)+\nonumber\\
    &+&\left[a_{12}e^{i\left(p_1k+p_2j\right)}+a_{21}e^{i\left(p_1j+p_2k\right)}\right]\vartheta(k-j)
  \end{eqnarray}
where $\vartheta(\centerdot)$ is the Heavyside function with $\vartheta(0)=1/2$. In the following, it is useful to write the eigenfunction \eqref{BA} as
\begin{equation}\label{BAXx}
   \phi_{jk}=e^{iPX}\left(a_{12}e^{ip\left|x\right|}+a_{21}e^{-ip\left|x\right|}\right)
  \end{equation}
where we have defined the centre-of-mass and relative coordinates of the two particles  by 
  \begin{equation}
   \left\{\begin{array}{l}
	X=\frac{j+k}{2}\\
    x=j-k
   \end{array}\right.
  \end{equation}
and their corresponding momenta by
  \begin{equation}
   \left\{\begin{array}{l}
    P=p_1+p_2\\
    p=\frac{p_1-p_2 \phantom{\frac{}{}}}{2}
   \end{array}\right.
  \end{equation}

 Thus, eq.ns \ref{EShIT} are satisfied if the energy $E$ is given by:
  \begin{equation}
  \label{Esc}
   E=-2J\left(\cos p_1+\cos p_2\right)=-4J\cos\frac{P}{2}\cos p
  \end{equation}
while:
  \begin{equation}
 y(P,p) \equiv   \frac{a_{21}}{a_{12}}=-\frac{U-4iJ\cos\frac{P}{2}\sin p}{U+4iJ\cos\frac{P}{2}\sin p} 
  \end{equation}
Notice that $y(P,p)$ has unitary module. By now imposing periodic boundary conditions ($  \phi_{j,-\frac{L}{2}}=\phi_{j,\frac{L}{2}+1}$), we can fix the values of the momenta $P$ and $p$, which have to satisfy the equations:
  \begin{equation}
   P_n=\frac{2\pi n}{L+1}
  \end{equation}
  \begin{equation}\label{eqsc}
   (-1)^ne^{ip(L+1)}=y\left(P_n,p\right)
  \end{equation}
with $n\in\{-L/2,\cdots,L/2\}$. Eq. \ref{eqsc} has to be solved numerically. For each value of $n$, we find $L/2+1$ solutions. The total number of independent eigenstates is therefore $(L+1)(L+2)/2$, as required for a two-boson system. The solutions for the relative momentum $p$ are of two types: real, leading to  {\it scattering} eigeinstates, and pure imaginary, leading to {\it bound} eigenstates. We will denote them with $\left|s\right>$ and $\left|b\right>$ respectively, with coefficients $\phi_{jk}^s$, $\phi_{jk}^b$ and energies  $E_s$ and $E_b$.

\begin{figure}[!t]
\centering
 \includegraphics[width=0.5\textwidth-9.5pt]{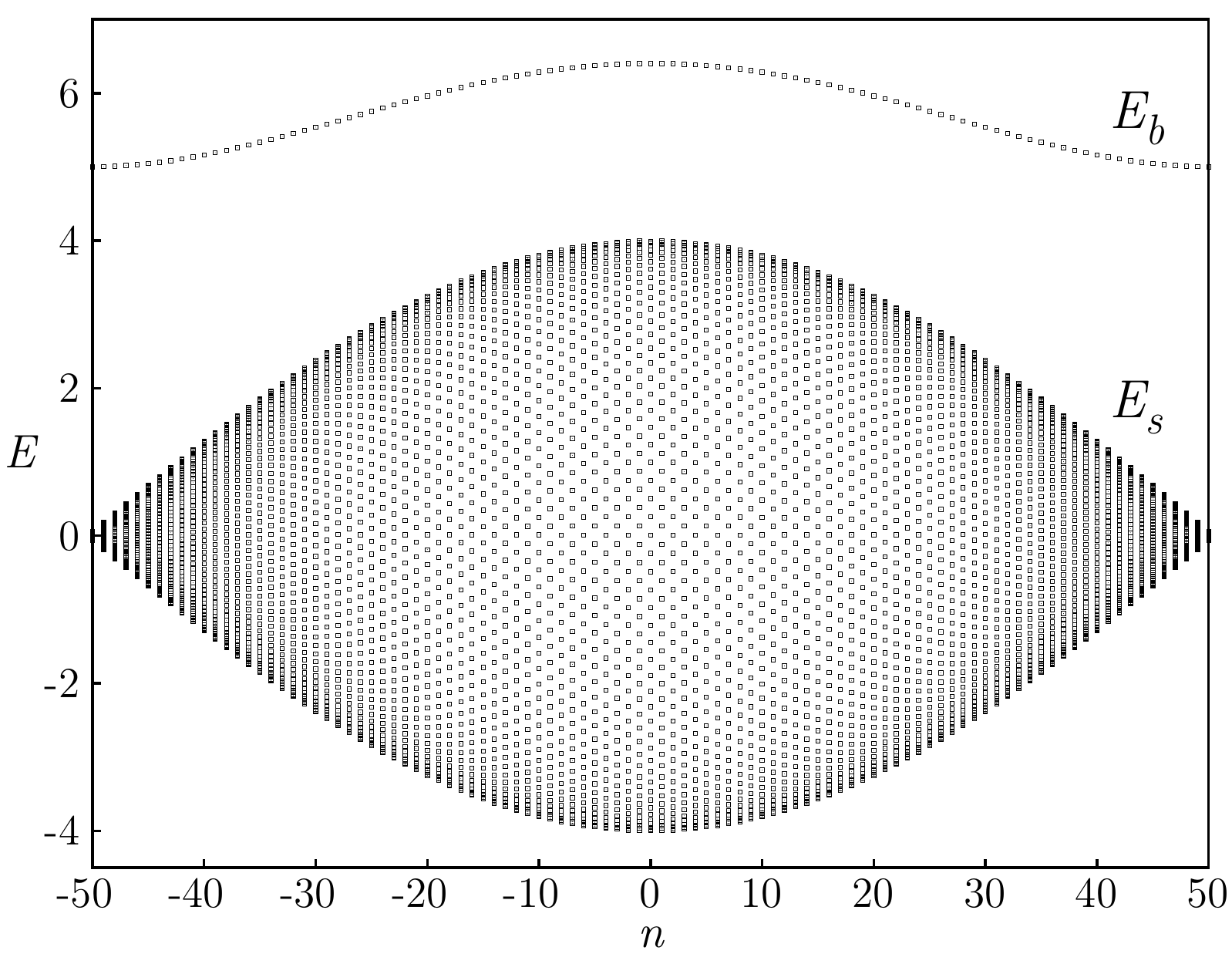}
 \caption{Energy spectrum of Eq. \ref{BHH} with $N=2$, $L+1=101$, $J=1$ and $U=5$.}\label{energies}
\end{figure}

As an example, in Fig. \ref{energies} we represent the full energy spectrum for $L+1=101$ and $U=5$: the scattering states form a band, becoming continuous in the thermodynamic limit; bound states exist even if the potential is repulsive  but their energies lie above the band.
It is easily seen from Eq. \eqref{eqsc}, that the whole spectrum is inverted by changing $U\rightarrow-U$. Therefore, in the attractive case, the bound states lie below the scattering ones, as expected. 
Let us observe that the energies of the bound states are shifted by changing $U$; in particular, if $|U|<4J$, the bound states close to the boundaries of the Brillouin zone, have energies lower than the top of the scattering band. 
These results fit perfectly with those obtained by Valiente and Petrosyan with different methods \cite{Valiente}.

We can now easily implement the time evolution as follows: if we choose the system to be, at $t=0$, in the initial state
  \begin{equation}\label{psi0}
   \left|\psi_0\right>=\sum_{j=-L/2}^{L/2}\psi^0_{jk}b_j^\dagger b_k^\dagger\left|0\right>
  \end{equation}
  the evolved state at time $t$ will be given by
  \begin{eqnarray}
   \left|\psi(t)\right>&=&e^{-it\hat{H}(U)}\left|\psi_0\right> \nonumber \\
 &=&\sum_sC_0^se^{-itE_s}\left|s\right>+\sum_bC_0^be^{-itE_b}\left|b\right>
  \end{eqnarray}
  where
  \begin{equation}
   C_0^{s/b}=\left<s/b\right|\left.\psi_0\right>=2\sum_{j,k=-L/2}^{L/2}\left(\phi_{jk}^{s/b}\right)^*\psi^0_{jk}
  \end{equation}
  We are interested in evaluating expectation values of observables on the evolved state. In particular, we focus on the density
   \begin{equation}
   \rho_j(t)=\frac{n_j(t)}{2}=\left<\psi(t)\right|\frac{\hat{n}_j}{2}\left|\psi(t)\right>
   \end{equation}
   and on the single and double occupations
   \begin{eqnarray}
    s_j(t) & = & \left<\psi(t)\right|\hat{n}_j\left(2-\hat{n}_j\right)\left|\psi(t)\right>\label{s}\\
    d_j(t) & = & \left<\psi(t)\right|\frac{\hat{n}_j\left(\hat{n}_j-1\right)}{2}\left|\psi(t)\right>\label{d}
   \end{eqnarray}
To compute them we need the matrix elements of $\hat{n}_j$ and $\hat{n_j}^2$ between the eigenstates of $H$, which are given by:
   \begin{eqnarray}
    \left<\alpha\right|\hat{n}_j\left|\beta\right> & = & 4\sum_{k=-L/2}^{L/2}\left(\phi_{jk}^\alpha\right)^*\phi_{jk}^\beta\\
    \left<\alpha\right|\hat{n}_j^2\left|\beta\right> & = & 4\left(\phi^\alpha_{jj}\right)^*\phi^\beta_{jj}+\left<\alpha\right|\hat{n}_j\left|\beta\right>
   \end{eqnarray}
   with $\alpha$, $\beta=s$, $b$.


\section{Dynamics of product states}\label{product}

In this Section, we will consider the time evolution, according to the Hamiltonian \eqref{BHH}, of an initial {\it product} state (in real space) either of the form:
\begin{equation}
\left|\psi_{PS}^{1}\right>  =  \frac{1}{\sqrt{2}}\left(b_0^\dagger\right)^2\left|0\right>
\end{equation}
corresponding to two bosons on the central site of the chain, or of the form:
\begin{equation}
\left|\psi_{PS}^{2}\right> =  b_0^\dagger b_1^\dagger\left|0\right>
\end{equation}
corresponding to a state with two bosons on two adjacent sites. The latter resembles the one of Ref. \onlinecite{Ronzheimer2013}, where product states of $N$ bosons on $N$ adjacent sites were considered. In the following we will consider just non-negative values of $U\in[0,31]$, since it was proven in Ref. \onlinecite{Schneider2012} that, for initial product states and for the observables we are considering, the dynamics at $\pm U$ is specular (see also Appendix \ref{theorem}).

The numerical results refer to a system of total size $L+1=25$, evolved up to a final time $t=4$ (from now on, we will measure times in units of $\hbar/J$ and choose $J=1$). They were obtained using the exact formulas of Section \ref{BAS}. 

\subsection{$\left|\psi_{PS}^{2}\right>$: two bosons on adjacent sites}

\begin{figure}[t]
\centering
 \includegraphics[width=0.5\textwidth-9pt]{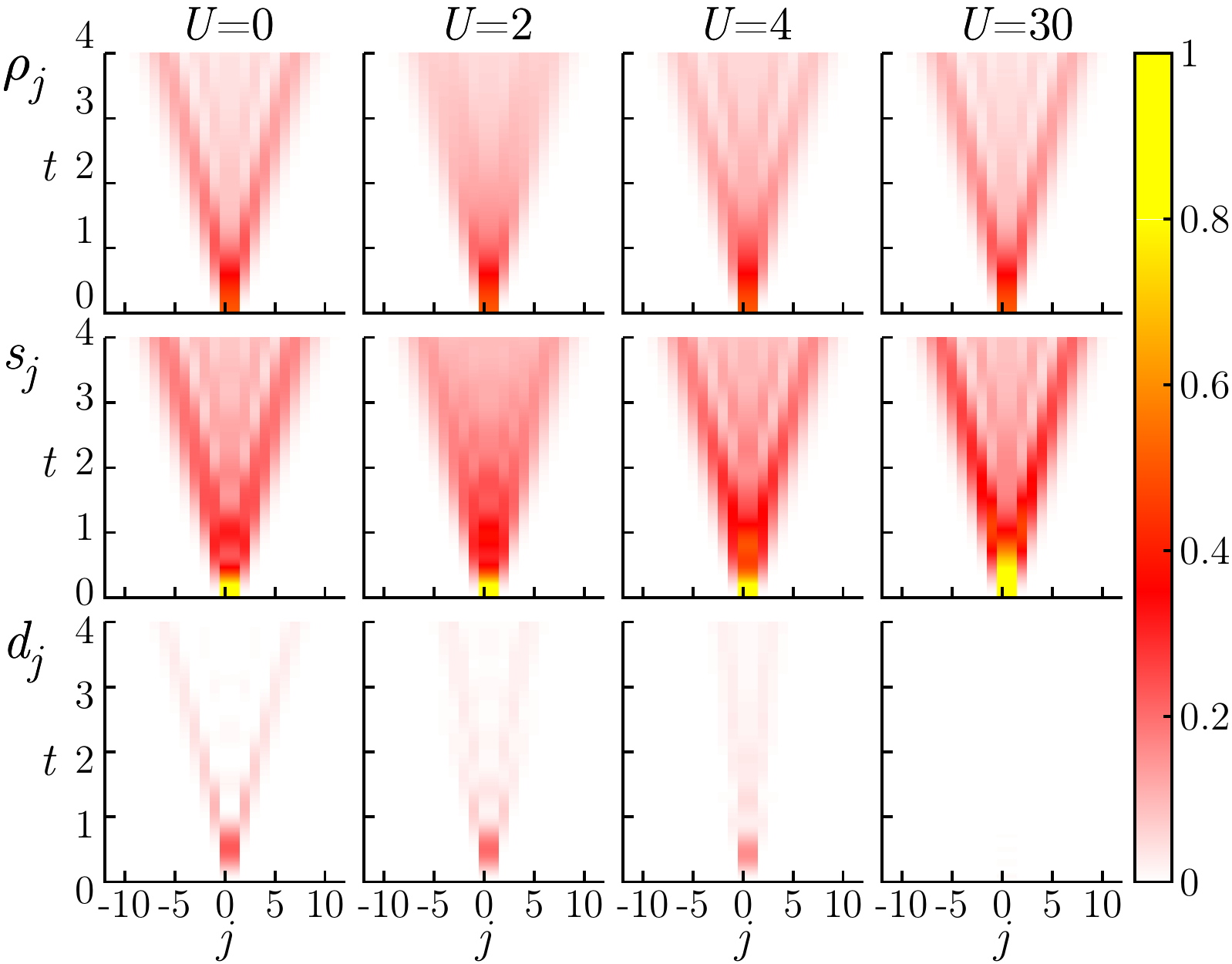}
 \caption{(Color online) Dynamical profiles of density $\rho_j$ (first row), single $s_j$ (second row) and double $d_j$ occupations (third row) for an initial product state $\left|\psi_{PS}^{2}\right>$ for $U=0$, 2, 4 and 30.}\label{occupationfirstkind}
\end{figure}

In Fig. \ref{occupationfirstkind} we show the dynamical profiles of $\rho_j(t)$, $s_j(t)$ and $d_j(t)$  at different values of the interaction. Starting from $\left|\psi_{PS}^{2}\right>$, the behaviour turns out to be quite peculiar: increasing $U$, the density profile, displaying a typical free form for $U=0$, changes as a result of the interaction and then tends to become free again at large $U$. This is explained by the fact that, for large $U$, the bosons tend to become {\it hard-core} \cite{Ronzheimer2013}, i.e., equivalent to free fermions \cite{Cazalilla2011}. This fact is possible just because the initial state is completely free from double occupations; this will not be the case for the initial states in the following Sections.
\begin{figure*}[t]
\includegraphics[width=1\textwidth]{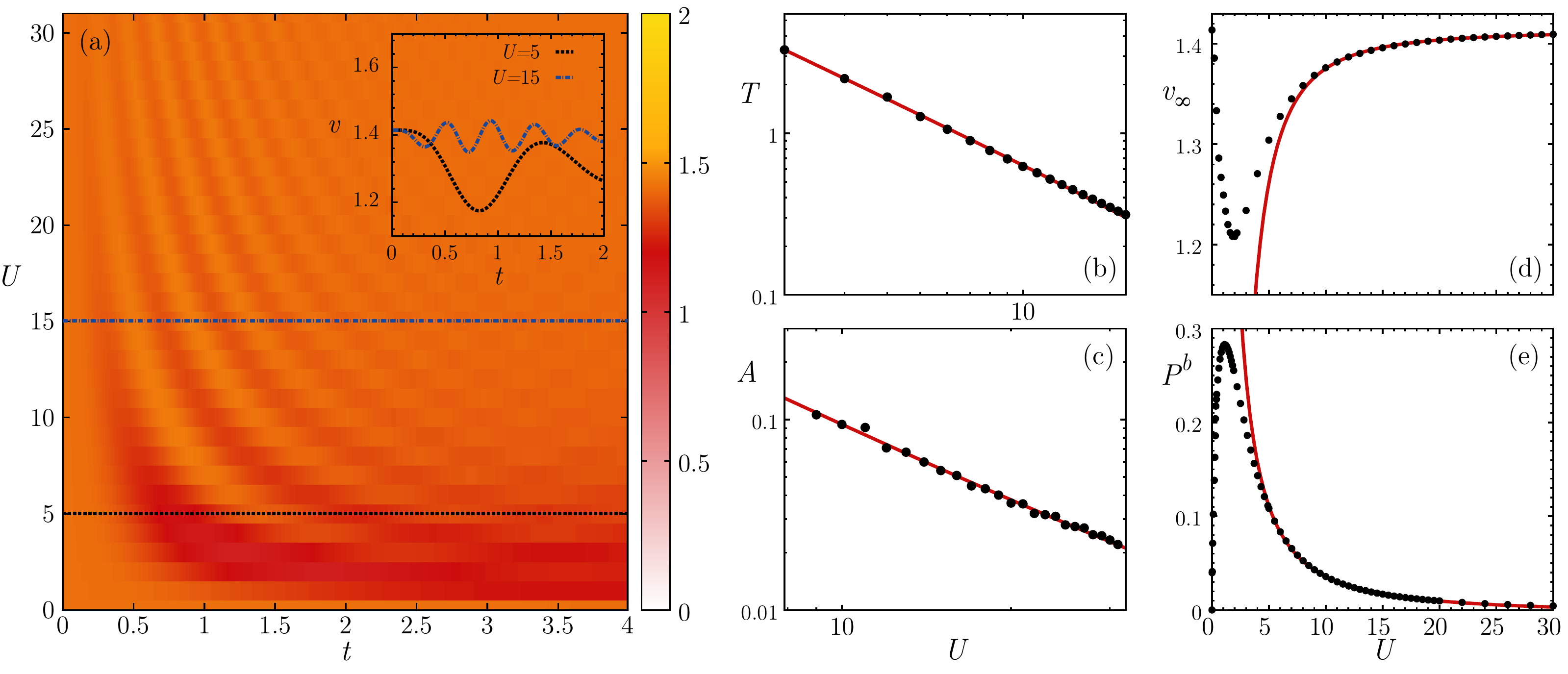}
\caption{(Color online) Features of the expansion velocity for an initial state $\left|\psi_{PS}^{2}\right>$. Panel (a): expansion velocity $v(t,U)$ for $U\in[0,31]$; inset: $v(t)$ for $U=5$ and $U=15$. Panels (b) and (c): large-$U$ dependence of $T(U)$ and $A(U)$, as obtained by fitting $v(t,U)$ according to Eq. \ref{fitform} (black dots: numerical data; red line: best fit, according to the formula $T(U)/A(U)=a_0/U^{a_1}$, giving $a_0\simeq6.41$, $a_1\simeq1.01$ for $T(U)$ and $a_0\simeq1.91$, $a_1\simeq1.31$ for $A(U)$). Panel (d): asymptotic expansion velocity $v_\infty(U)$ (black dots: numerical data; red line: best fit according to the formula $v_\infty(U)\sim\sqrt{2}+a/U^2$, giving $a\simeq-3.80$). Panel (e): projection $P^b(U)$ of the evolved state on the bound states of \eqref{BHH} (black dots: numerical data; red line: best fit according to the power law $P^b(U)=a_0/U^{a_1}$, giving $a_0\simeq1.33$, $a_1\simeq1.52$).}
\label{2bosons_L24_MI_l2_Pb-U}
\end{figure*}
A similar behaviour is visible in the single-occupation profile, while the double-occupation one differs significantly: in the large-$U$ regime, double occupations are almost absent, as a sign of the fermionization of the dynamics. Moreover, one can see that, for intermediate values of the interaction (e.g. $U\simeq 5$), the double occupation of the central sites are quite stable, significantly vanishing only at large times. This phenomenon is known as {\it quantum distillation} \cite{HM2009}, and was observed for many particle wave functions, both in the case of spin-$1/2$ fermions \cite{HM2009,Bolech2012} and of bosons \cite{Ronzheimer2013}. 

We analyze now the expansion velocities, as defined by Eq. \ref{expv}. Fig. \ref{2bosons_L24_MI_l2_Pb-U}(a) shows a contour plot of $v(t,U)$. We note that: i) the initial velocity $v(t=0)$ is independent of $U$; ii)  as shown also in the inset, for $U\neq 0$, damped oscillations of $v$ as a function of $t$ appear, with a period $T(U)$ that decreases as $U$ increases (we will come back to this point in the next paragraph; see equation \eqref{fitform}). 
The asymptotic expansion velocities for large time at each $U$, that we call $v_{ \infty}(U)$, can also be estimated, by fitting $v(t)$ in the range $[t^*,4]$, with $t^*=2$ (this allows to neglect the transient oscillatory region), with the formula:
\begin{equation}\label{fitform}
v(t)=v_{ \infty}(U)+A(U)\cos\left(\frac{2\pi t}{T(U)}+\phi\right)/t^\eta
\end{equation}

The shape of $v_{ \infty}(U)$ for $N=2$ is reported in Fig. \ref{2bosons_L24_MI_l2_Pb-U}(d) and is very similar to the one observed in the many-body case \cite{Ronzheimer2013}: $v_{ \infty}(U)$ approaches the value $\sqrt{2}$ in the two limit cases $U=0,\;+\infty$ (see Appendix \ref{freevel}) and displaying a minimum for an intermediate value $U\simeq2$. The large-$U$ behaviour of $v_\infty(U)$ has been studied, in the many-body case, in a recent work \cite{Sorg2014}: the authors were able to show that, for the current initial state and in the strong-repulsive regime, $v_\infty(U)\sim\sqrt{2}+a/U^2$, were $a$ is a constant. We check this prediction by our numerical data: the result is shown in Fig. \ref{2bosons_L24_MI_l2_Pb-U}(d). The agreement between numerics and the analytical prediction is excellent.

The period $T(U)$ and  the amplitude $A(U)$ of the damped oscillations of $v(t)$ display a clean power-law behaviour for large-enough $U$ (see Figs. \ref{2bosons_L24_MI_l2_Pb-U}(b) and (c)). In particular, $T(U\geq5)$ decays with an exponent that is very close to 1 (see the caption of Fig. \ref{2bosons_L24_MI_l2_Pb-U}). 

We can understand the behaviour of the asymptotic velocity by considering the role of bound states during the evolution. Let us consider the (time-independent) projection of the wave functions on the subspace spanned by the bound states:
\begin{equation}\label{Pbeq}
 P^b(t)=\sum_b\left|\left<b\right|\left.\psi(t)\right>\right|^2 \equiv P^b(0) =\sum_b\left|C_0^b\right|^2
\end{equation}
In panel (e) of Fig. \ref{2bosons_L24_MI_l2_Pb-U} we plot $P^b$ as a function of $U$. A comparison with the panel (d) shows that the larger the projection $P^b$, the smaller the rate of expansion of the wave packet is: if the projection of the wave function over the bound states is small, then the evolution is free-like, and the expansion is fast; otherwise, when the initial state has a a large projection on bound states, the expansion velocity decreases (as shown in Fig. \ref{2bosons_L24_MI_l2_Pb-U}(e), for large $U$, $P^b(U)$ is well approximated by a power law: see the caption of Fig. \ref{2bosons_L24_MI_l2_Pb-U}). Despite the qualitative agreement between the shapes of $v_\infty$ and $P^b$ as functions of $U$, there is a quantitative displacement in the positions of the maximum of $P^b$ with respect to the minimum of $v_\infty$: this effect can be partially explained by the difficulty of extracting the true value of $v_\infty$ in the small-$U$ regime from the numerical data (on the contrary, $P^b$ is, in any case, an exact quantity). 

In conclusion, we have shown that bound states play a key role for the expansion dynamics of the system. We further strengthen this interpretation by considering different cases in the following sections.

\subsection{$\left|\psi_{PS}^{1}\right>$: two bosons on the same site}

\begin{figure}[t]\centering
 \includegraphics[width=0.5\textwidth-8pt]{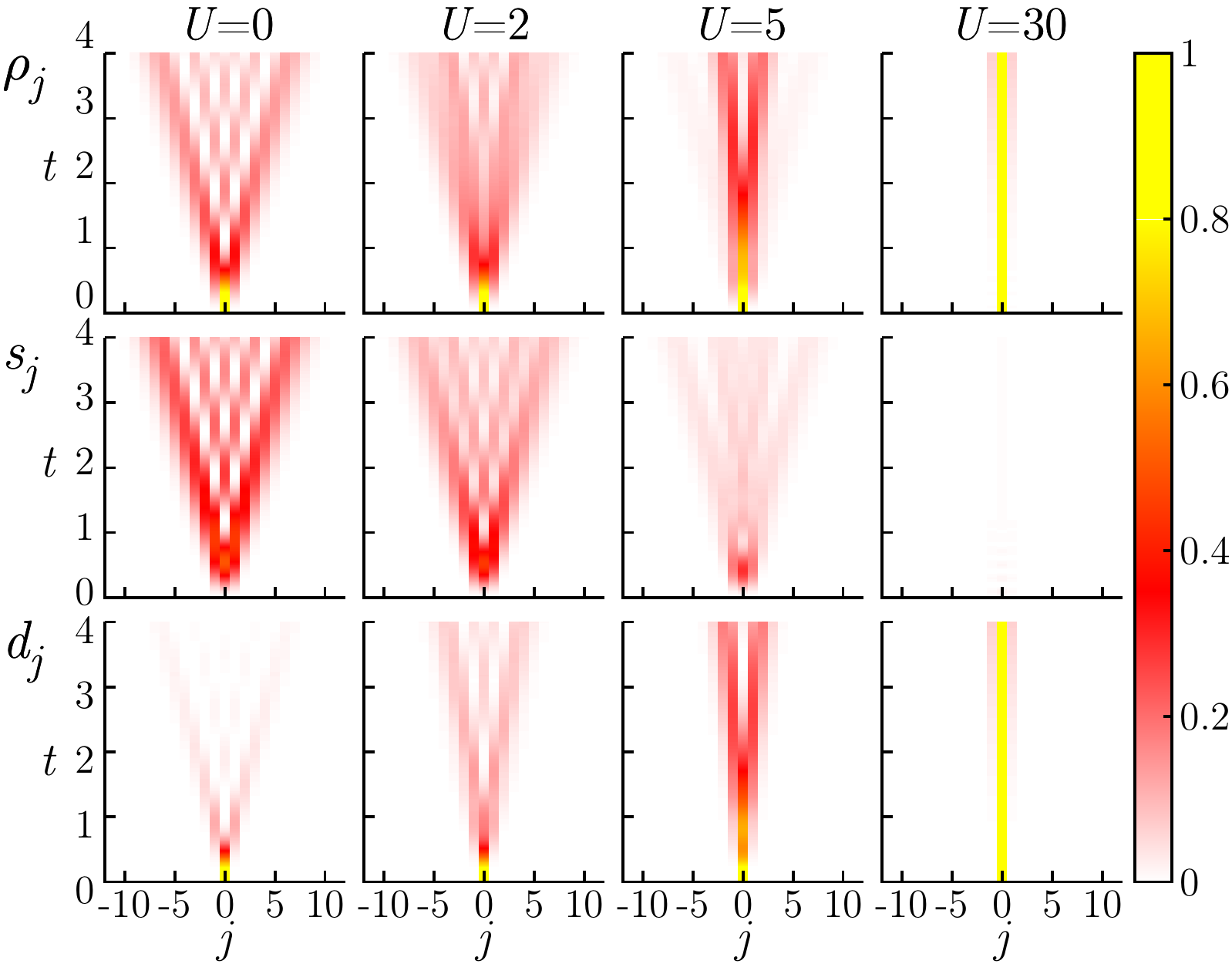}
 \caption{(Color online) Dynamical profiles of density $\rho_j$ (first row), single $s_j$ (second row) and double $d_j$ occupations (third row) for an initial product state $\left|\psi_{PS}^{1}\right>$ for $U=0$, 2, 5 and 30.}\label{occupationsecondkind}
\end{figure}

We now analyze the dynamics for an initial product state $\left|\psi_{PS}^{1}\right>$ in which two bosons are confined on the same site. We remark that the expansion of a generalization of this state, i.e., a Mott-insulator like state with double occupancy, was studied in Ref. \onlinecite{Muth2012}. The dynamical profiles of the density, single and double occupations are shown in Fig. \ref{occupationsecondkind}. 
Again, for small $U$ the expansion is free-like, but a free regime is no longer approached at $U=+\infty$. This feature is explained theoretically by the fact that the mapping from hard-core bosons to free fermions can be applied only if the double occupations are (almost) absent in the initial state, a condition which is clearly not true for $\left|\psi_{PS}^{1}\right>$ \cite{Ronzheimer2013}. The evolution of the double occupations for this initial state is stable also for large $U$ while the single occupations, zero at $t=0$, are finite only in the small $U$ regime. Here we notice that the role of double and single occupations is reversed compared to the $\left|\psi_{PS}^{2}\right>$ case.

\begin{figure*}[t]
 \includegraphics[width=\textwidth]{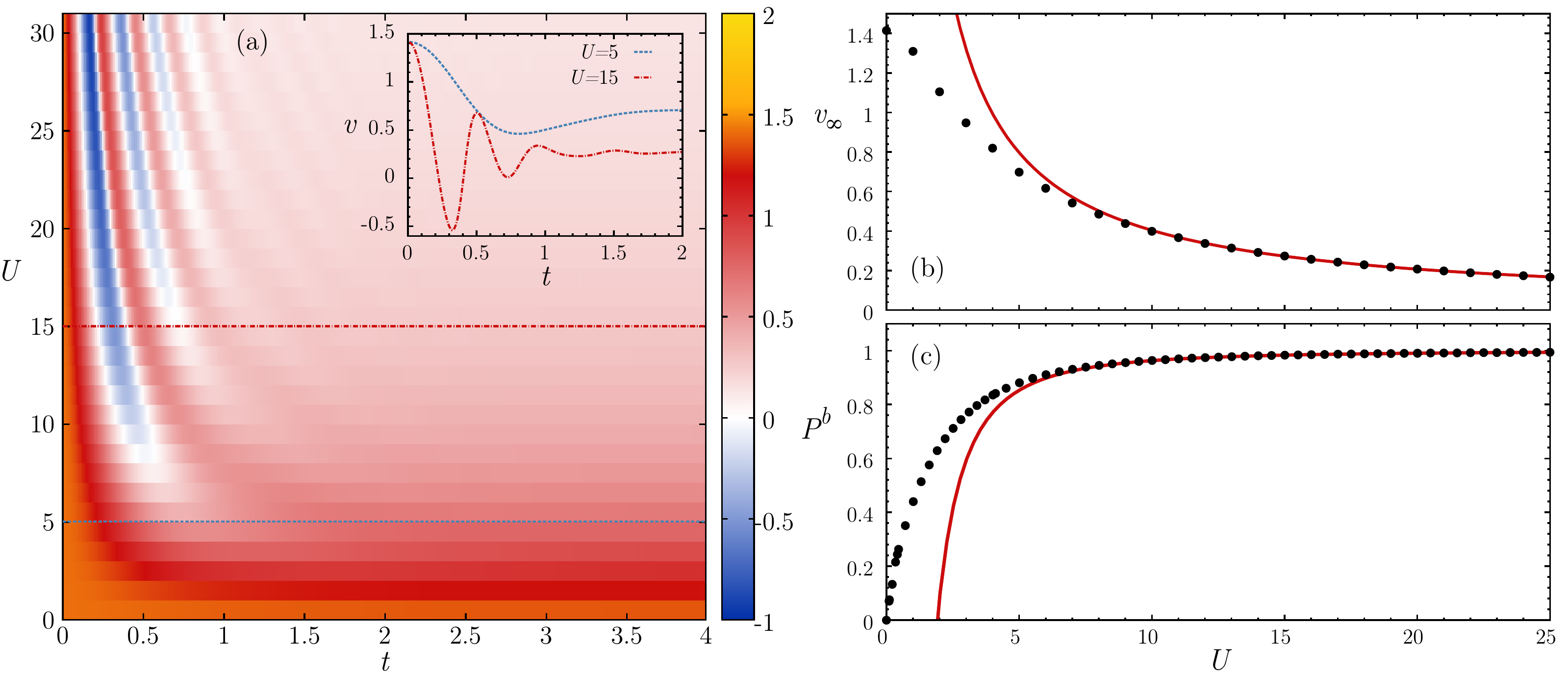}%
\caption{(Color online) Features of the expansion velocity for an initial state $\left|\psi_{PS}^{1}\right>$. Panel (a): expansion velocity $v(t,U)$ for $U\in[0,31]$; inset: $v(t)$ for $U=5$ and $U=15$. Panel (b): asymptotic expansion velocity $v_\infty(U)$ (black dots: numerical data; red line: best fit according to the formula $v_\infty(U)\sim a_0/U^{a_1}$, giving $a_0\simeq3.84$, $a_1\simeq0.97$). Panel (c): projection $P^b(U)$ of the evolved state on the bound states of \eqref{BHH} (black dots: numerical data; red line: best fit according to the power law $P^b(U)=1+a/U^2$, giving $a\simeq-3.67$).}
\label{2bosons_L24_MI_l=1_v-U}
\end{figure*}

In Fig. \ref{2bosons_L24_MI_l=1_v-U}(a) we show the contour plot of $v(t,U)$.  
As in the previous  case, $v(t=0)$ is independent from $U$ and, at  a given $U$, oscillations of $v$ as a function of $t$ are present, with decreasing period as $U$ increases. We notice that now, at large $U$ and early times, the velocity periodically assumes negative values, signaling a "breathing" behaviour of the wave packet at these times. We then study the asymptotic velocity $v_{\infty}(U)$, by fitting $v(t,U)$ in the interval $t\in[t^*,4]$, with $t^*=3$, just with a constant, since oscillation are highly suppressed. The results are plotted in Fig. \ref{2bosons_L24_MI_l=1_v-U}(b): we see that $v$, as a function of $U$, presents a monotonically decreasing behaviour for all the considered values of the interaction. Moreover, in this case, the large-$U$ values of $v$ are well fitted by a power law $v(U)\sim a_0/U^{a_1}$, with $a_1\simeq1$ (see the caption of Fig. \ref{2bosons_L24_MI_l=1_v-U}).

Even in this case, an explanation of the shape of $v(U)$ in terms of the projection over the bound states can be given; $P^b$ as a function of $U$ is shown in Fig. \ref{2bosons_L24_MI_l=1_v-U}(c). Large velocity and free expansion correspond to a small projection of the wave function on the bound states, and {\it viceversa}, illustrating the importance of the bound states even in this situation. Moreover, for large $U$, the projection on the bound states is shown to saturate to $1$ as $1+1/U^2$ (see Fig. \ref{2bosons_L24_MI_l=1_v-U}(c) and its caption).

\section{Dynamics of entangled states}\label{entangled}

In this section, we consider a different kind of initial states, that we call {\it entangled}, since they do not result from a direct product in real space, but from the ground state of a non-interacting Bose-Hubbard Hamiltonian in an open box of length $l$; in the two-particle sector this yields:
 \begin{equation}\label{SF0}
    \left|\psi_{ES}^l\right>=\frac{1}{\sqrt{2}}\left(\tilde{b}_1^\dagger\right)^2\left|0\right>
   \end{equation}
 where, $\tilde{b}_1^\dagger$ represents the operator creating a particle of minimum momentum (see, e.g., Ref. \onlinecite{Taddia_Phd}). For this state, $s^0_j$ and $d^0_j$ are simultaneously non-zero. Initial states different from products in real space were also considered in Ref. \onlinecite{Vidmar2013}. Our results were obtained using the exact methods of Secs. \ref{BAS}. 

\begin{figure}\hspace{-1.2em}
 \includegraphics[width=0.5\textwidth]{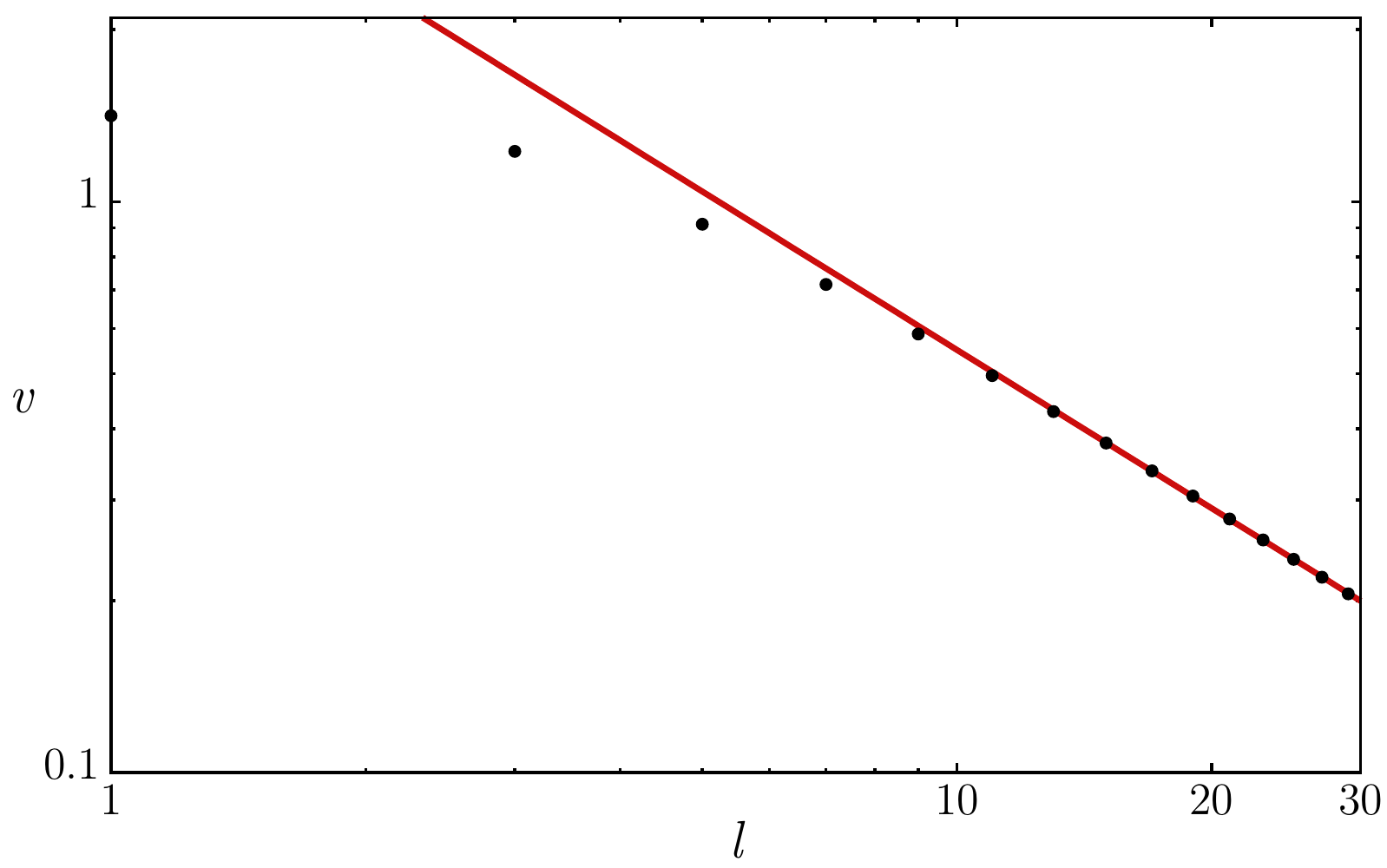}
 \caption{(Color online) Expansion velocity of an entangled wave packet in the non-interacting case as a function of the initial confinement length $11\leq l\leq31$, with a total system size $L+1=51$. The black dots are the numerical data, while the red solid line is the best fit for large $l$, performed with the formula $y=a_0/x^{a_1}$, resulting in $a_0\simeq4.37$, $a_1\simeq0.91$. The plot is in log-log scale.}\label{SF-U0-v-l}
\end{figure}

We first studied the evolution of the entangled state with a non-interacting Hamiltonian. As in the product cases of Sec. \ref{product}, the free evolution separates into two beams: in Sec. \ref{lattice} we will show that this effect is due to the lattice structure, and disappears when the size of the initial wave packet is much larger than the lattice spacing. The time dependence of $v$ in the non-interacting case is trivial too, $v(t)$ being a constant. We notice that in the $l=2$ case the expansion velocity is, as shown in Fig. \ref{SF-U0-v-l}, exactly the one of a product state of two neighbour particles, i.e., $\sqrt{2}$. Instead, increasing $l$, the expansion velocity decreases; its $l$-dependence, for $11\leq l\leq31$ and $L+1=51$, is also shown in Fig. \ref{SF-U0-v-l}. As stated in the caption, the dependence on $l$ is compatible with a power law $\sim l^{-1}$, which  is the behaviour one would have in the continuum case for a Gaussian wave packet (see Eq. \ref{sigma-t}), where $v\sim\sigma^{-1}$.

\begin{figure*}[!!!h!!!t]
 \includegraphics[width=\textwidth]{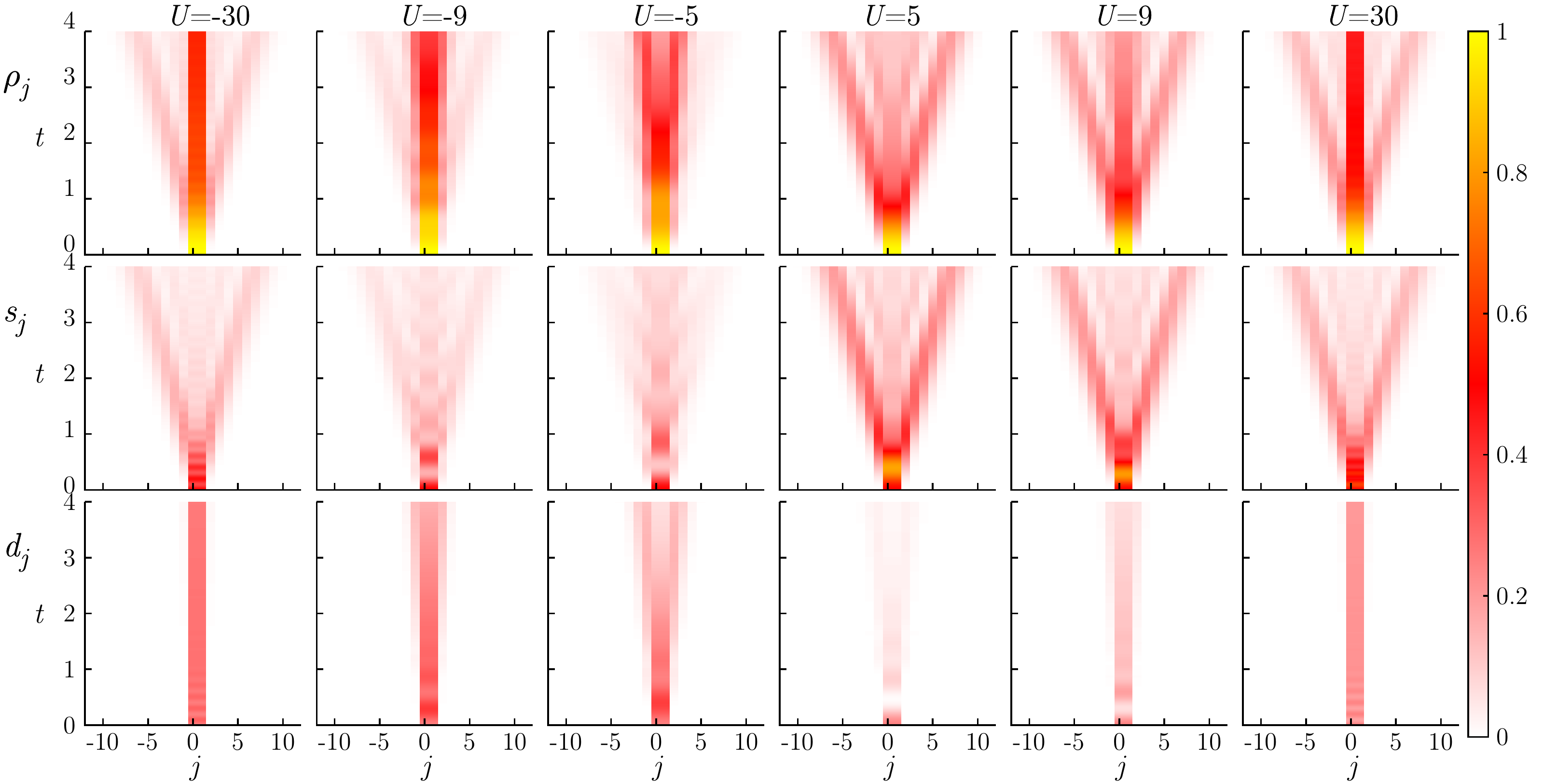}
 \caption{(Color online) Dynamical profiles of density $\rho_j$ (first row), single $s_j$ (second row) and double $d_j$ occupations (third row) for an initial entangled state $\left|\psi_{ES}^{2}\right>$ for $U=-30$, -9, -5, 5, 9 and 30.}\label{SF-profiles-l2}
\end{figure*}

We then analyze the situations with $U\neq0$. Now, the initial state does  no longer satisfies the conditions for the validity of the $U\leftrightarrow-U$ theorem (see Appendix \ref{theorem}), so we have to perform different calculations for positive and negative $U$. 
The profiles of density and single and double occupations are depicted in Fig. \ref{SF-profiles-l2} for $l=2$ and some significant values of $U\in[-31,31]$ (the profiles for $l=3$, 4 display just quantitative differences).

For all the considered values of $l$, the profiles at large-positive and large-negative $U$ are very similar to each other, while some differences are encountered at smaller $|U|$'s: for small negative $U$'s, the behaviour is similar to the large $|U|$'s ones, while for small positive $U$'s it is more similar to the free ones. As we shall see, these facts are explained, even in this case, by considering the projection of the wave function on the bound states, $P^b(U)$ (Eq. \ref{Pbeq}). Moreover, in the $l=2$ case (but not in the others), $v(t=0)$ is seen to be independent of $U$ and equal to $\sqrt{2}$, in analogy with the product-state cases.

\begin{figure}[!ht]
 \includegraphics[width=0.5\textwidth]{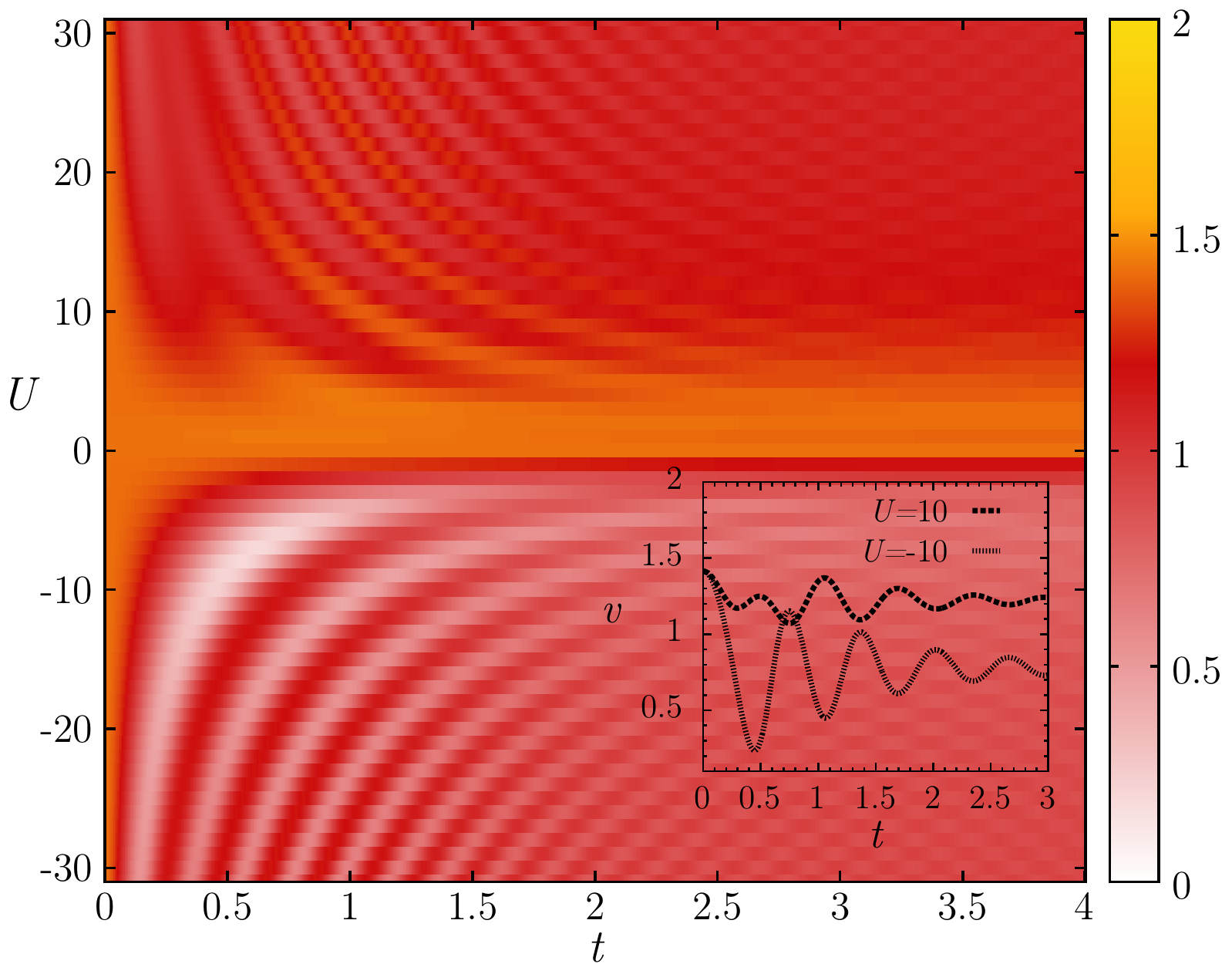}
 \caption{(Color online) Expansion velocity $v(t,U)$ for an initial entangled state $\left|\psi_{ES}^2\right>$ and $U\in[-31,31]$; inset: $v(t)$ for $U=\pm10$.}\label{SF-profiles-l2-contour}
\end{figure}

We then compute the expansion velocities, and we plot them as functions of $t$ and $U$, in Fig. \ref{SF-profiles-l2-contour}.
The $v(t,U)$'s for the three considered values of $l$ are quite similar to each other: in any case, $v$ oscillates in a damped way as a function of $t$, even if the amplitude decreases as $l$ grows, marking the fact that, for large $l$, the bosons are diluted enough to propagate in a nearly free way.

\begin{figure*}[!ht]
 \includegraphics[width=\textwidth]{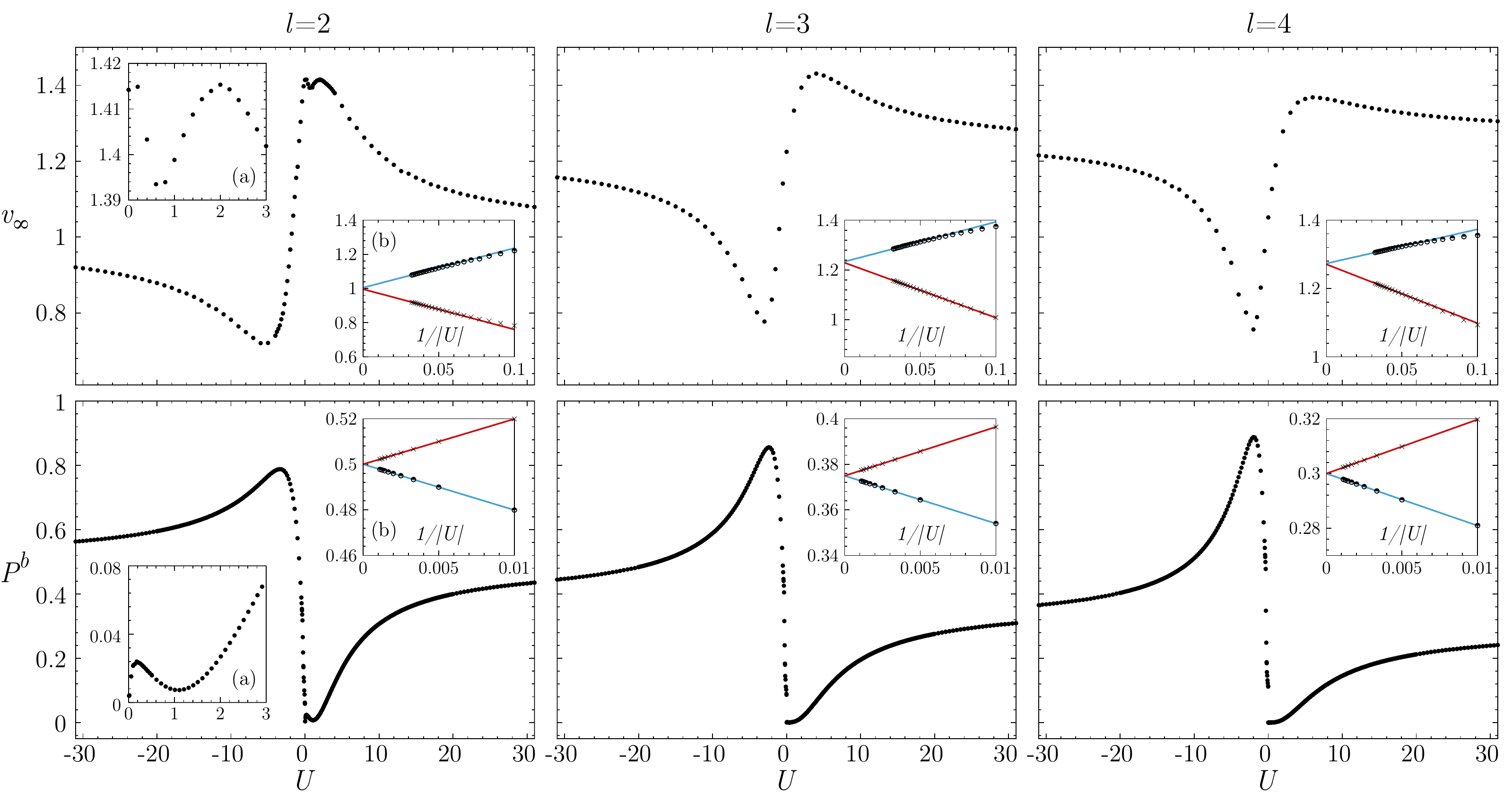}
 \caption{(Color online) First row: asymptotic expansion velocities as functions of $U\in[-31,31]$ for initial entangled states with $l=2$, 3 and 4. Second row: projection of the wave function over the bound states as a function of $U\in[-31,31]$ for initial entangled states with $l=2$, 3 and 4. Insets (a) of the $l=2$ data: magnification of the regions $U\in[0,3]$; remaining insets:  best fits of the large-$|U|$ data with power laws $a_0+a_1/|U|$ (circles/crosses: numerical data with $U>/<0$; red/light-blue solid line: best fit for $U>/<0$). Results of the best fits (in any case the $a_0$ for the positive and negative $U$'s coincide to the second digit): first row: left panel: $a_0\simeq1.00$; central panel: $a_0\simeq1.23$; right panel: $a_0\simeq1.27$; second row: left panel: $a_0\simeq0.50$; central panel: $a_0\simeq0.37$; right panel: $a_0\simeq0.30$.}\label{Pb-U-ent}
\end{figure*}

The analysis performed in order to extract $v_\infty(U)$ at large times is quite different from the product cases. In particular, for $l=2$ and 3 we choose $t^*=2$, while, for $l=4$, we choose $t^*=3$; moreover, for $l=2$ and for negative $U$'s of the $l=3$ case the fit of $v(t)$ is performed by the 9-parameters formula $v(t)=v_\infty(U)+A_1\cos\left(\frac{2\pi t}{T_1}+\phi_1\right)/t^{\eta_1}+A_2\cos\left(\frac{2\pi t}{T_2}+\phi_2\right)/t^{\eta_2}$, while for $l=4$ and the positive $U$'s of the $l=3$ the oscillations are so damped that a fit by means of a constant is enough. Following this procedure, we are able to extract reliable values for $v_\infty(U)$, but not for the remaining fit parameters. In the first row of Fig. \ref{Pb-U-ent} we report $v_\infty(U)$ for $l=2,3$ and $4$.

The three shapes display some common features: in all cases, an absolute maximum and an absolute minimum are present; moreover, before the minimum and after the maximum $v_\infty(U)$ is monotonic, tending, for $U\rightarrow\pm\infty$, to definite asymptotic values. We estimate them using a simple power law $v_\infty(U)\sim a+b/U$, as shown in the insets of Fig. \ref{Pb-U-ent}. Very remarkably, the asymptotic values at $\pm\infty$ are the same in any case; in particular, for $l=2$, they are very close to 1: this fact can be easily understood by looking at the occupation profiles in Fig. \ref{SF-profiles-l2}, showing that both for negative and positive $U$ the evolution is dominated by the coherent propagtion of double occupations. Another interesting aspect of such curves is the fact that the asymptotic value of $v_\infty$ is closer to the value of the maximum as $l$ is increased, reflecting the fact that the initial wave packet is more dilute with increasing $l$. Moreover, the $l=2$ curve displays a second maximum at $U=0$, and the corresponding value is $v(0)=\sqrt{2}$, as previously remarked.

We then compute the quantity $P^b(U)$, defined in Eq. (\ref{Pbeq}) as the projection of the wave function on the bound states. In the second row of Fig. \ref{Pb-U-ent} we plot this quantity {\it vs} $U$: in analogy with the product case, larger velocities correspond to smaller $P^b$'s, and {\it viceversa}, and a direct correspondence between velocities and projection on the bound states is established. One of the most remarkable features of these plots is, as shown in the inset (a) of the left panel of the second row of Fig. \ref{Pb-U-ent}, the small positive $U$'s behaviour for $l=2$: while $v_\infty(U)$ displays, in this regime, a two-maxima shape, $P^b(U)$ possesses, in correspondence, two minima: this fact is a strong evidence to the correctness of our interpretation. However, in the same region, $P^b$ is not able to capture the quantitative details of $v_\infty(U)$, such as, for instance, the precise position of the two maxima in the $l=2$ case: this can be partially explained, as for the $\left|\psi_{PS}^2\right>$ case, by the difficulty of extracting reliable values of $v_\infty$ in the small-U regime. Finally, we extrapolate the asymptotic values of $P^b$, that are seen to coincide for $U\rightarrow\pm\infty$. In particular, for $l=2$, $P^b(\pm\infty)$ is very close to 1/2. Increasing $l$, this value is seen to decrease, reflecting the less dense nature of the initial wave packet.


\section{Effect of the lattice on the expansion dynamics}\label{lattice}

 As already emphasized, one of the most remarkable aspects related to the expansion of quantum matter on a lattice is the appearance of two quasi-coherent beams, departing from the centre of the initial wave packet. This effect is at odd with what happens in the continuum, where a Gaussian wave-packet remains Gaussian if the evolution is free, with a spread $\sigma(t)$ growing with time (see, e.g., Ref. \onlinecite{Messiah1999}). Such beams have been observed, apart from the present work, theoretically and experimentally, both for bosons \cite{QuasiCond, Ronzheimer2013,Vidmar2013} and for fermions \cite{Langer2012,Vidmar2013}, and seem to be absent in the continuum case \cite{Schneider2012}. The effects of the discreteness have already been stressed by Rigol and collaborators \cite{QuasiCond}, who argued that the most populated states in the diffusion process are at a well-defined value of the momentum $p_0$, depending on the mean value of the energy $\left<H\right>$ (with $p_0=\pm\pi/2$ when $\left<H\right>=0$). This argument supports the fact that the two-beam effect scarsely depends on the choice of the initial state.

\begin{figure*}[th]
 \includegraphics[width=\textwidth]{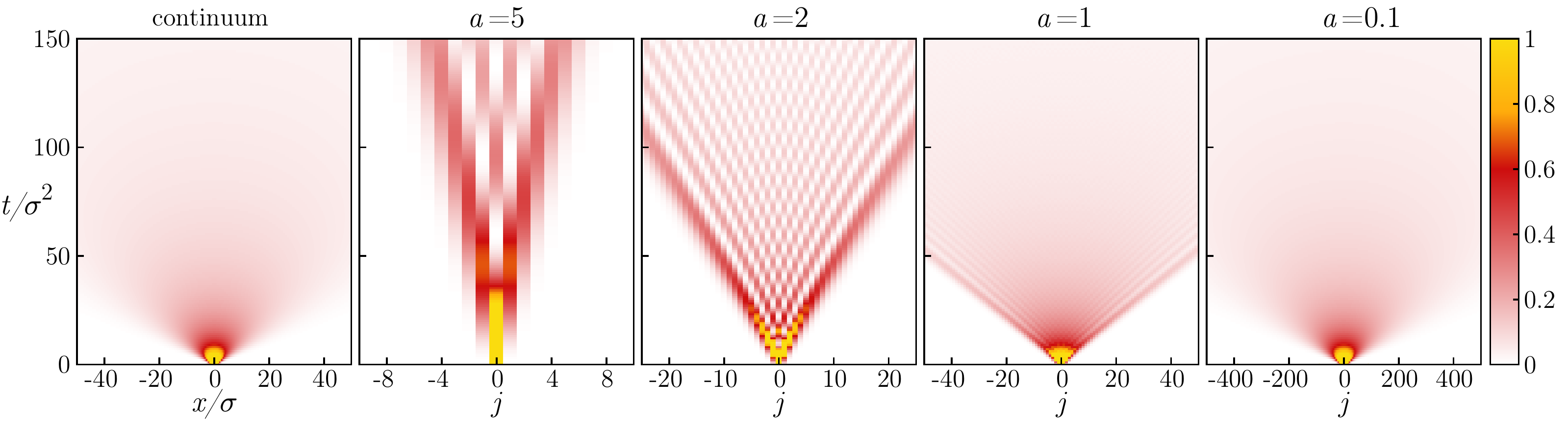}
 \caption{(Color online) Expansion of a Gaussian wave packet with $\sigma=0.1$ in the continuum (a) and in the lattice, with lattice spacing $a=5\sigma$ (b), $2\sigma$ (c), $\sigma$ (d) and $0.1\sigma$ (e). We restricted the plot to the region $x/\sigma$, $ja\in[-50,50]$.}\label{latcont}
\end{figure*}

Let us briefly recall the situation in the continuum case, where the spatial coordinate $x\in\mathbb{R}$ \cite{Messiah1999}. If we start with a Gaussian wave packet of standard deviation $\sigma$, the particle density has a Gaussian shape at any time $t$,  
\begin{equation}
     \rho(x,t)=\frac{1}{\sigma(t)\sqrt{\pi}}e^{-\frac{x^2\phantom{\frac{}{}}}{\sigma^2(t)}}
\end{equation}
with a time-dependent variance:
\begin{equation}\label{sigma-t}
     \sigma^2(t)=\sigma^2+\left(\frac{\hbar t}{m\sigma}\right)^2
\end{equation}
as shown in Fig. \ref{latcont}(a) for $\sigma=0.1$ (we put $m=\hbar=1$, space is measured in units of $\sigma$ and time in units of $m\sigma^2/\hbar =\sigma^2$).

On an infinite lattice, the single particle Hamiltonian is
    \begin{equation}
     \hat{H}=-\frac{\hbar^2}{2ma^2}\sum_{j\in\mathbb{Z}}\left(b_j^\dagger b_{j+1}\phantom{^\dagger} + b_{j+1}^\dagger b_j\phantom{^\dagger}\right)
    \end{equation}
  where the constant in front is chosen to ensure the correct continuum limit (see below). Its orthonormalized eigenstates and eigenenergies are given respectively by:
    \begin{equation}
     \left|p\right>=\sqrt{\frac{a}{2\pi}} \sum_{j\in\mathbb{Z}}  e^{ipja}  \, b_j^\dagger\left|0\right>
    \end{equation}
\begin{equation}
     E(p)=-\frac{\hbar^2}{ma^2}\cos(pa)
    \end{equation}
 As in the continuum, we choose an initial state with a normalized Gaussian profile:
    \begin{equation}
     \left|\psi(0)\right>=A_{\sigma,a} \sum_{j\in\mathbb{Z}}   e^{-\frac{j^2a^2\phantom{\frac{}{}}}{2\sigma^2}}   \, b_j^\dagger\left|0\right>
    \end{equation}
 with $A_{\sigma,a}=\left[\theta_3\left(0,e^{-\frac{a^2\phantom{\frac{}{}}}{\sigma^2}}\right)\right]^{-1/2}$,  $\theta_3(z,q)$ being the third elliptic theta function \cite{Gradshteyn2007}. Straightforward calculations yield: 
    \begin{equation}
     \left|\psi(t)\right>=\sum_{j\in\mathbb{Z}}\psi_j(t)\,  b_j^\dagger\left|0\right>
    \end{equation}
    with
    \begin{equation}
     \psi_j(t)=\frac{\int_{-\pi}^\pi dp\theta_3\left(-\frac{p}{2},e^{-\frac{a^2\phantom{\frac{}{}}}{2\sigma^2}}\right)e^{i\frac{\hbar t\phantom{\frac{}{}}}{ma^2}\cos p}e^{ijp}}{2\pi\sqrt{\theta_3\left(0,e^{-\frac{a^2\phantom{\frac{}{}}}{\sigma^2}}\right)}}
    \end{equation}

 The dynamical profile of the density can then be computed from $n_j(t)=\left|\psi_j(t)\right|^2/a$, 
 where the lattice spacing has been put in the denominator in order to recover the correct dimension, for the comparison with the continuum case. The numerical results are shown in Fig. \ref{latcont}(b)-(e) for $\sigma=0.1$ and decreasing values of $a$. For large $a$ the presence of the two beams, departing from the centre of the initial wave packet, is manifest. Moreover, it is evident that, decreasing the value of the lattice spacing at fixed $\sigma$, the profile density becomes similar to that in the continuum. This is confirmed quantitatively by comparing the values of the densities in the continuum and for $a=0.1\sigma$ (the difference being $10^{-3}$ or less). We have therefore shown that the two matter beams, already present in the single-particle case, are a consequence of the presence of the lattice.


\section{Conclusions and outlooks}\label{conclusions}

The present paper deals with the dynamical effects of bosonic pairs in a 1D lattice, which may be coupled in bound states by short range interactions, both in the attractive and repulsive case. In the latter situation, the bound states (at most one for each value of the centre-of-mass momentum) originate from the discrete structure of the linear-combination-of-atomic-orbitals lattice, and the resulting energy levels lie \emph{above} the band of the scattering states. The two-particles case exploited here can be approached exactly, by means of the Bethe ansatz. This makes it possible to identify the bound states effects unambiguosly, by studying how the two-particles state evolves in time, depending on its initial projection on the bound states themselves. In particular, we have studied the single, double and total occupation probability of the evolving state, and the resulting expansion velocity, for two classes of initial conditions: a) \emph{product states}, corresponding to bosons with well defined initial positions in real space, and b) \emph{entangled states}, corresponding to the ground states of a non interacting Bose-Hubbard Hamiltonian, in an open box of size $l$. In the case a), a symmetry argument (not appliable to the case b)) ensures the invariance of the results for attractive and repulsive interactions. In general, the results fit well with physical intuition: the larger the initial projection on the bound states, the larger the double occupation probability, and the smaller the expansion velocity. In addition, a number of quantitative behaviours were found, especially in the large-$|U|$ cases. This provides a quantitative support to the importance of the bound pair states, a point that might escape one's attention, or even look unphysical in the \emph{repulsive} case. Moreover, we discussed the role of the lattice in the shape of the evolved wave function, showing that it is the responsible of the separation of the evolved packet in two wave fronts propagating in opposite directions.

Remarkably, for the $\left|\psi_{PS}^2\right>$ initial wave packet, that was already considered in a previous work both experimentally and theoretically \cite{Ronzheimer2013}, we found that the two-body case displays several qualitative and quantitative analogies with the many-body one, meaning that the main features of the expansion are present already in the few-body situation \cite{Kessler2013}. It is therefore quite likely that the effects we studied are relevant in the many-particles case too and can be identified by suitable projection methods. This aspect is, indeed, the next development we plan for future researches, in addition to the extension to different kind of initial states \cite{Vidmar2013}, interacting Fermi systems \cite{Schneider2012,Sorg2014} and the consideration of equilibration/thermalization issues \cite{Goldstein2014}. We therefore believe that our study can open the way to a number of further investigations about the role of bound states in the dynamics of many-body systems.

\acknowledgments

We thank L. Barbiero, M. Dalmonte and D. Vodola for useful discussions. L. T. acknowledges financial support from IP SIQS.


\appendix

\section{Expansion velocities for product states in the non-interacting case}\label{freevel}

In this Appendix, we show explicitly that, in the non-interacting case, the expansion velocity, as defined by Eq. \ref{expv}, must be given by \cite{Ronzheimer2013}
\begin{equation}\label{freevelformula}
 v=\frac{\sqrt{2}Ja}{\hbar}
\end{equation}
We will set, in the following, $\hbar=a=J=1$.

Since we are in the non-interacting case, we can just consider one-particle wave packets that we assume to have initially the generic form 
\begin{equation}
 \left|\psi^0\right>=\frac{1}{\sqrt{L+1}}\sum_p c_p\, \tilde{b}_p^\dagger\left|0\right>
\end{equation}
being $\tilde{b}_p^\dagger$ the Fourier transform of $b_j^\dagger$. The density profile $n_j(t)$ can be written as a Fourier transform, in the $L\rightarrow\infty$ limit:
\begin{equation}
 n_j(t)=\frac{1}{2\pi}\int_{-\pi}^\pi dqe^{-iqj}f_q(t)
\end{equation}
where
\begin{equation}
 f_q(t)=\frac{1}{2\pi}\int_{-\pi}^\pi dp \, c_p^* c_{p+q}\, e^{i\left[\epsilon(p+q)-\epsilon(p)\right]t}f_p(t)
\end{equation}
$\epsilon(p)$ being the one-particle energies.

Let the initial wave packet be centered around the central site $j=0$. Then, as in standard probability theory, the second momentum of the occupation distribution, i.e., $R^2(t)$ (see \eqref{Rsquare}), can be computed from the generating function $f_q(t)=f(q,t)$ by the relation
\begin{equation}
 R^2(t)=-\frac{\partial^2f(q,t)}{\partial q^2}
\end{equation}
When we take the initial state corresponding to one particle in the central site $j=0$, i.e. $c_p=1$, it is easy to see that
\begin{equation}
 R^2(t)=2J^2t^2
\end{equation}
implying Eq. (\ref{freevelformula}), as a consequence of the definition (\ref{expv}).

\section{$U\leftrightarrow-U$ inversion theorem}\label{theorem}

In this Appendix, we state the theorem, first proven in Ref. \onlinecite{Schneider2012} (see also Refs. \cite{Fradkin1991,Valiente2010}), ensuring the invariance of the dynamical expectation values of certain operators for a Hubbard-like Hamiltonian and a proper initial wave packet. We then prove that entangled states, as defined in Sec. \ref{entangled}, do not satisfy the hypothesis of the theorem.

Before stating it, we have to define the time-reversal operator $\hat{R}_t$ as
    \begin{equation}
     \hat{R}_te^{-it\hat{H}}\hat{R}_t^\dagger=e^{it\hat{H}}
    \end{equation}
    and the $\pi$-boost operator $\hat{B}_\pi$ as
    \begin{equation}
     \hat{B}_\pi b_j\hat{B}_\pi=e^{i\pi j}b_j
    \end{equation}
    The theorem then states that if an observable quantity $\hat{O}$ is invariant under the actions of the above defined operators, and the state at the initial time of evolution $\left|\psi^0\right>$ is time-reversal invariant and just acquires a phase under the action of $\hat{B}_\pi$, then $\left<\psi(t)\right|\hat{O}\left|\psi(t)\right>$ is the same if the time evolution is ruled by a Bose-Hubbard Hamiltonian with $\pm U$. The proof was carried out for the fermionic Hubbard Hamiltonian, but it can be trivially extended to the bosonic case.

It was proven in Ref. \onlinecite{Schneider2012} that the product states we consider in Sec. \ref{product} satisfy the hypothesis of the theorem; on the contrary, the entangled states of Sec. \ref{entangled} do not. Indeed,
    \begin{eqnarray}
      \hat{B}_\pi\left|\psi^0\right>&=&\hat{B}_\pi\frac{1}{\sqrt{2}}\left(\tilde{b}_1^\dagger\right)^2\left|0\right>=\nonumber\\
      &=&\frac{1}{\sqrt{2}}\left[\sqrt\frac{2}{l+1}\sum_{j=1}^l\sin\left(p_1j\right)\hat{B}_\pi b_j^\dagger\hat{B}_\pi\right]^2\left|0\right>=\nonumber\\
      &=&\frac{1}{\sqrt{2}}\left[\sqrt\frac{2}{l+1}\sum_{j=1}^l(-1)^j\sin\left(p_1j\right)b_j^\dagger\right]^2\left|0\right>
    \end{eqnarray}
    that manifestly does not differ from $\left|\psi^0\right>$ just by a phase factor. Therefore, in the entangled case, we must also consider negative $U$'s.




\end{document}